\documentclass[prl,aps,twocolumn,nofootinbib]{revtex4}

\newcommand{\bear}{\begin{array}}  \newcommand{\eear}{\end{array}}
\newcommand{\bea}{\begin{eqnarray}}  \newcommand{\eea}{\end{eqnarray}}
\newcommand{\beq}{\begin{equation}}  \newcommand{\eeq}{\end{equation}}
\newcommand{\bef}{\begin{figure}}  \newcommand{\eef}{\end{figure}}
\newcommand{\bec}{\begin{center}}  \newcommand{\eec}{\end{center}}
\newcommand{\non}{\nonumber}

\def\EQ#1{Eq.~(\ref{#1})}

\begin{document}

\title{Smallness of Baryon Asymmetry from Split Supersymmetry}

\author{Shinta Kasuya$^a$ and Fuminobu Takahashi$^b$}

\affiliation{
$^a$ Department of Information Science,
     Kanagawa University, Kanagawa 259-1293, Japan\\
$^b$ Institute for Cosmic Ray Research,
     University of Tokyo, Chiba 277-8582, Japan}

\date{January, 2005}

\begin{abstract}
The smallness of the baryon asymmetry in our universe
is one of the greatest mysteries and may originate from some profound
physics beyond the standard model.
We investigate the Affleck-Dine baryogenesis in split supersymmetry, 
and find that the smallness of the baryon asymmetry is directly
related to the hierarchy between the supersymmetry breaking squark/slepton 
masses and the weak scale. Put simply, the baryon asymmetry is small
because of the split mass spectrum.
\end{abstract}

\pacs{nnn}

\maketitle

\setcounter{footnote}{1}
\renewcommand{\thefootnote}{\fnsymbol{footnote}}

{\it Introduction}.---
The hierarchies in the energy scales pose naturalness issues. 
Cosmological constant is found to be of order
(meV)$^4$ \cite{CC}, while the fundamental scale should be the Planck scale (or
the weak scale). To keep the higgs mass around $O(100)$ GeV, a new physics
such as supersymmetry must appear at $O(1)$ TeV,  although  precision electroweak measurements 
have pushed up this scale higher than $O(5-7)$ TeV, threatening supersymmetry as a solution
to naturalness problem associated with the hierarchy problem \cite{Higgs}.

These naturalness issues may not be a problem any more in the context of the
anthropic landscape of string theory \cite{Susskind:2003kw}: we may live in
such a vacuum that has a very small cosmological constant {\it and} a 
very small higgs mass compared with fundamental scale. The fine tuning of
the latter led to split supersymmetry (SUSY) framework
\cite{split}, where the SUSY breaking scale in the
visible sector is liberated from the weak scale.

The idea of split SUSY \cite{split} is to abandon solving 
the hierarchy problem in the standard model (SM) and allow all the scalars except
for a higgs to get a heavy mass. Then we can avoid many
problems in supersymmetric SM (SSM), such as the absence of a light higgs
and most of the sparticles, too fast decay of the proton by the
dimension five operators, SUSY flavor and CP problems, and the
cosmological gravitino and moduli problems.

The scale of the SUSY breaking is arbitrary in general in split SUSY.
The two guiding principles usually taken are the gauge coupling
unification at the grand unified theory scale and the lightest
SUSY particle as dark matter. In order to satisfy both
requirements, gauginos should be kept as light as TeV scale. 
The simplest mass spectrum is such that all the squarks and
sleptons are at $\tilde{m}$ presumably much higher than
the weak scale, while keeping all the gauginos and higgs very light. In this case, 
the gaugino masses $m_{1/2}$ should be suppressed by some
mechanism \cite{Arkani-Hamed:2004yi}, which also keeps $A$-term 
small: $A \sim m_{1/2} \sim {\rm TeV}$. 

How to create the baryon asymmetry of the universe is one of the
biggest issues in the modern cosmology. 
 Through the observations of light nuclei \cite{BBN} and cosmic microwave
background anisotropies \cite{WMAP}, its 
abundance is known to be very small: the
baryon-to-entropy ratio should be around $10^{-10}$. In this Letter,
we show that the split SUSY beautifully explains the smallness of the
baryon asymmetry of the universe.

In the split SUSY scheme, many aspects of baryogenesis are
significantly altered compared to usual supersymmetric SM. For
instance, the electroweak baryogenesis \cite{Kuzmin:1985mm} does not work 
both because the strongly first order phase transition cannot be attained
due to large stop mass and because available CP violation is also very
limited for split mass spectrum. Therefore, other baryogenesis
mechanisms should account for the baryon asymmetry of the universe. Here we
concentrate on the Affleck-Dine (AD) baryogenesis~\cite{Affleck:1984fy}.

The AD mechanism utilizes the squark/slepton condensate to generate
the baryon asymmetry. Since the squark/slepton masses $\tilde{m}$ are
much larger than the weak scale in split SUSY, the resultant asymmetry
is suppressed compared to that in the SSM. 
As we will see below, in the simplest case, the baryon-to-entropy ratio
is given by the ratio between the two split mass scales, the weak scale
and the heavy scalar masses $\tilde{m}$: $n_B/s \sim 0.1 A/\tilde{m}$, 
which predicts $\tilde{m} \sim O(10^{12})$ GeV for $n_B/s \sim
10^{-10}$ and $A \sim$ TeV. Thus the smallness of the present baryon
asymmetry can be ascribed to the split mass spectrum.

{\it Affleck-Dine mechanism and split SUSY}.---
In the minimal supersymmetric standard model (MSSM), flat directions
including squarks and/or sleptons have baryon
and/or lepton numbers, so it can be considered as the
AD field, $\Phi$. They acquire  heavy SUSY breaking masses of order
$\tilde{m}$. In addition, let us assume that the AD field
is lifted by a non-renormalizable operator, $W=\Phi^n/n M^{n-3}$,
where $M$ is a cut-off scale. 
In order to produce baryon asymmetry, the AD field must have a
torque, whose force comes from the baryon-number-violating $A$-term. 
As mentioned above, the same mechanism for obtaining the light gaugino masses
makes $A$ as small as $m_{1/2}$. The relevant scalar potential for $\Phi$ is 
\beq
V(\Phi) = \tilde{m}^2 |\Phi|^2 + \left(\frac{A\Phi^n}{n M^{n-3}} 
+ {\rm h.c.} \right) + \frac{|\Phi|^{2n-2}}{M^{2n-6}}.
\eeq

During inflation supergravity and K\"ahler couplings with the
inflaton lead to a Hubble-induced mass term, 
\beq
\delta V = c H_I^2 |\Phi|^2,
\eeq
where $c$ is a coefficient of order unity and $H_I$ is the Hubble
parameter during inflation. If $c H_I^2 + \tilde{m}^2< 0$, this
negative mass makes the origin unstable and sets the 
initial value for $\Phi$ away from the origin: 
\beq
\label{eq:min}
|\Phi_{\rm inf}| \simeq {\rm min}\left[ \left(H_I
    M^{n-3}\right)^{\frac{1}{n-2}}, M_p \right],
\eeq
where $M_p=2.4\times 10^{18}$ GeV is the reduced Planck mass, and
this upper bound reflects the fact that the scalar potential becomes
exponentially steep above $|\Phi| \sim M_p$. It is worthy of note
that $H_I$ should be larger than $\tilde{m}$ in order to develop the
large initial amplitude of $\Phi$. In the usual case of low-scale soft
SUSY breaking mass, $\tilde{m} \sim {\rm TeV}$, this condition is
satisfied for most inflation models. However, if $\tilde{m}$ is much
larger than the TeV scale as in the split SUSY scheme, this condition
sets a nontrivial and severe constraint on the energy scale for the 
inflation. We will come back to this issue below.

After inflation the Hubble parameter starts to decrease. The AD field
tracks the instantaneous minimum given by \EQ{eq:min} with $H_I$
replaced by $H$, the Hubble parameter at that time. When 
$H \sim \tilde{m}$, the AD field comes to oscillate, and the baryon
number density is efficiently produced at the same time: 
\beq
n_B \simeq B_\Phi \delta  \,A  |\Phi_{\rm osc}|^2 ,
\eeq
where $B_\Phi$ is the baryon charge of $\Phi$,  $\delta \sim O(0.1)$ 
represents a CP phase, and $|\Phi_{\rm osc}| \equiv (\tilde{m}
M^{n-3})^{1/(n-2)}$ is assumed to be smaller than $M_p$. The resultant
baryon-to-entropy ratio depends on the subsequent thermal history. For
simplicity, let us first consider the case that the AD field dominates
the energy density of the universe when it decays. Then the
baryon-to-entropy ratio is 
\beq
\label{eq:nbs}
\frac{n_B}{s} \simeq \frac{n_B}{\rho_{tot}/T_d}
\sim \frac{B_\Phi \delta \,A  T_d}{\tilde{m}^2}
\sim 0.1\frac{m_{1/2}}{\tilde{m}},
\eeq
where $T_d \sim \tilde{m}$ is the decay temperature of the AD field,
which decays into the gauginos and quarks/leptons.
Note that the decay does not proceed until  the effective masses
of decay particles become smaller than the parent mass,
$\tilde{m}$, resulting in $T_d\sim \tilde{m}$. 

From Eq.~(\ref{eq:nbs}) we can see that the baryon asymmetry is
suppressed by the hierarchy between the gaugino mass and the
squark/slepton mass. Because of the large hierarchy in split SUSY, a
desired value of $n_B/s \sim 10^{-10}$ can be obtained if $\tilde{m}$
is around $10^{12}$ GeV for $m_{1/2} = 1$ TeV.
This is rather amazing result; the observed
small baryon asymmetry is simply determined by the split mass spectrum
in the SSM, independent of $n$ or $M$ as long as the AD field
dominates the universe. This should be contrasted to the usual case of
$\tilde{m} \sim$ TeV  leading to $n_B/s \sim O(0.1)$.

The conditions for the AD field to dominate the universe are easily
achieved. If the reheating is completed before the AD field starts
oscillating, i.e., $T_{RH} >\sqrt{ \tilde{m} M_p}$,
\beq
\label{ineq1}
|\Phi_{\rm osc}| > \left(\tilde{m} M_p^3 \right)^\frac{1}{4}
\eeq
must be satisfied, where $T_{RH}$ is the reheating temperature after
inflation. On the other hand, for 
$\sqrt{\tilde{m} M_p} > T_{RH} > \tilde{m}$, we obtain a similar
inequality, 
\beq
|\Phi_{\rm osc}| >\sqrt{\frac{\tilde{m}}{T_{RH}} } M_p.
\eeq
It is easy to see that (\ref{ineq1}) is satisfied for e.g., $n=6$ and
$M \gtrsim M_p$.

 It should be emphasized that 
cosmological gravitino problem \cite{gravitino} can be avoided, 
since the gravitino mass $m_{3/2}$ can be much larger than the TeV
scale in split SUSY. Thus, the reheating temperature can be as large
as $10^{16}$ GeV. However it is still nontrivial whether the
gravitinos dominate the universe and produce extra entropy at the
decay, destroying a simple relation \EQ{eq:nbs}. Let us therefore
clarify the condition for gravitinos not to dominate the universe. For
$m_{3/2} \gtrsim \tilde{m}$, the gravitinos are mainly produced
through scattering processes, although one can easily see the
gravitinos produced in this way give only negligible contribution to
the energy density of the universe. On the other hand, for 
$\tilde{m} > m_{3/2}$, potentially dangerous process is the decay of
the squarks/sleptons into gravitinos, and the abundance of the
gravitinos is solely determined by this process 
\cite{Arkani-Hamed:2004yi}:
\begin{equation}
    \label{gravitino}
    Y_{3/2} \sim 10^{-5} N
    \left(\frac{\tilde{m}}{10^{12} {\rm GeV}}\right)^3
    \left(\frac{m_{3/2}}{10^9 {\rm GeV}}\right)^{-2}
    \left(\frac{g_*(\tilde{m})}{10^2} \right)^{-3/2},
\end{equation}
where $g_*(\tilde{m})$ counts the relativistic degrees of freedom at
$T = \tilde{m}$, and $N$ is the number of thermally populated
squarks/sleptons at that time. Note that since the decay temperature
of the AD field is of the order of $\tilde{m}$, not all the squarks
and sleptons might be thermally populated. Also, even if none of them
are thermally produced, there is still contribution from the decay of
the AD field (squark/slepton condensate) into gravitinos. Thus $N$
varies from a few to a few tens, depending on the detailed structure of
the sfermion mass spectrum and composition of the flat direction. In
order to keep the direct relation between the baryon asymmetry of the
universe and the mass hierarchy in split SUSY, the gravitino should
not dominate the energy density of the universe when it decays. Thus,
we obtain the constraint on the gravitino mass as 
\bea
    m_{3/2} &\gtrsim& 10^9 N
    \left(\frac{g_*(\tilde{m})}{10^2} \right)^{-3/5}
    \left(\frac{g_*(T_{3/2})}{10^2} \right)^{1/10} \non\\
&& \times       \left(\frac{\tilde{m}}{10^{12} {\rm GeV}}\right)^{6/5}
    {\rm GeV}.
\eea
where $T_{3/2}$ is the decay temperature of the gravitinos:
\beq
T_{3/2} = 4 \times 10^3  
 \left(\frac{g_*(T_{3/2})}{10^2} \right)^{-1/4}
\left(\frac{m_{3/2}}{10^9 {\rm GeV}} \right)^{3/2} {\rm GeV}.
\eeq
When the gravitino mass is smaller than $10^9$ GeV, there is an extra
entropy production, which dilutes the baryon
asymmetry. The dilution factor is estimated as
$\sim (m_{3/2}/10^9$ GeV$)^{-5/2}$. 

In the above discussion, we have assumed that the $Q$ balls~\cite{qball} 
are not formed. $Q$ balls may delay the decay of the AD
field and affect the direct relation between the baryon asymmetry and
the mass hierarchy. Here we show that $Q$ balls are actually absent in the
context of split SUSY. In order to have the $Q$-ball configuration,
the effective potential of the AD field must be shallower than
$|\Phi|^2$. Including one-loop corrections to the squark/slepton
masses, the effective potential scales as $|\Phi|^{2+2K}$, where
$K$ is the coefficient of the one-loop corrections. 
Hence the $Q$-ball solution exists, if and only if $K$ is negative.
If squark/slepton masses are comparable to the gaugino masses, i.e.,
$\tilde{m} \sim m_{1/2}$, the one-loop correction is dominated by
negative contribution from the gauginos, therefore $Q$-ball configuration
exists, leading to $Q$-ball formation. However, if  
$\tilde{m} \gg m_{1/2}$ as in split SUSY, this is not necessarily the case.
In particular, for $\tilde{m}/m_{1/2} \sim 10^9$,  $K$ is always positive, 
irrespective of whether or not the flat direction includes the
third-generation fields. Therefore, there is no $Q$-ball configuration
in the split SUSY.

Now let us consider fluctuations inherited in the baryon asymmetry.
If the AD field acquires primordial fluctuations during inflation,
the baryonic isocurvature fluctuation generally arises. In particular,
as extensively studied in the context of curvaton scenario \cite{curvaton},
the primordial fluctuations in the AD field cannot account for the
present density fluctuations, since the baryonic isocurvature would
then become too large \cite{Hamaguchi:2003dc}. Thus the primordial
fluctuations in the AD field must be either subdominant compared to 
the inflaton-induced adiabatic fluctuations or suppressed by the
Hubble-induced mass term and/or A-term.

Next let us study what if the AD field decay before its domination of the
universe. In this case, the entropy of the  universe almost entirely comes 
from the inflaton, leading to
\beq
\label{eq:nbs2}
\frac{n_B}{s} \simeq \frac{ B_\Phi \delta \,A 
|\Phi_{\rm osc}|^2}{\tilde{m}^2 M_p^2} {\rm min}
\left[T_{RH}, \sqrt{\tilde{m} M_p}\right].
\eeq
Although the split mass hierarchy explains the smallness of the baryon
asymmetry to some extent,  the direct and simple relation like 
\EQ{eq:nbs} is lost. 

Finally, we would like to mention that in our scenario radiation
of the present universe mainly comes from the AD field rather than the
inflaton \cite{MSSM}.  Since the SM degrees of freedom are naturally created by
the decay of the AD field,  this feature can be considered as a merit of making 
easy to build inflation models. One of the other ways to realize the feature
is to drive inflation itself within the SSM.  For instance, 
the MSSM flat direction could drive the (chaotic)
inflation with the quartic potential~\cite{KMT} or even the quadratic
potential with the inflaton mass $m_I \sim \tilde{m} \sim 10^{13}$ GeV
in split SUSY.

{\it Conclusion}.---
We have considered the AD baryogenesis in split SUSY model,
in which all the scalars except for a finely-tuned higgs get a heavy mass
$\tilde{m}$, while gauginos (and higgsinos) have a weak scale mass.
It was found that the small baryon asymmetry of the universe directly
results from the mass hierarchy between them. 
To be specific, as is clear from \EQ{eq:nbs},
the heavy scalar mass scale $\tilde{m}$ should be around $10^{12}$ GeV,
to explain the baryon-to-entropy ratio $n_B/s \sim 10^{-10}$ for
$m_{1/2} \sim$ TeV. It is interesting that thus predicted value of
$\tilde{m}$ is close to the upper limit $\tilde{m} \lesssim 10^{13}$
GeV, which comes from the requirement that the lifetime of gluinos
should not exceed about $10^{16}$ sec~\cite{split}.  

In addition, this scenario suggests high reheating temperature of the
inflaton, which in turn implies high-scale inflation. Thus, the
detailed observation of the tensor mode or BB mode of CMB polarization 
may favor or refute our scenario.

{\it Acknowledgments.}---
  F.T.  would like to thank the Japan Society for
Promotion of Science for financial support. 



\end{document}